# Identification and prioritization of urban traffic bottlenecks

Nimrod Serok[1], Shlomo Havlin2, Efrat Blumenfeld Lieberthal1*
[1]Azrieli School of Architecture, Tel Aviv University, Tel Aviv 6997801, Israel
[2]Department of Physics, Bar-Ilan University, Ramat Gan 52900, Israel

**Abstract**

The increasing urbanization process we have been witnessing in the last decades is accompanied by a significant increase in traffic congestion in cities around the world. The effect of the congestion is represented in the enormous time people spent on roads leading to significant money waste and air pollution. Here, we present a new methodology for identification, cost evaluation, and thus, prioritization of congestion sources, the jam bottlenecks. It extends existing methods as it is based on network analysis of the entire road network and can be applied to different traffic models. Our results show that the macro-stability, presented by scaling characteristics of the traffic bottlenecks, overshadows the existence of meso-dynamics, where the bottlenecks change their location in time and space. This means that to plan and manage traffic jams in different locations and at different times, it is required to implement a framework, as developed here, that tracks traffic and evaluates the relative effect of each evolving bottleneck on the entire road network.

**Problem Definition**

The 21st century can be characterized as the century of the cities. Since 2008, more than 50% of the world's population lives in urban areas. The increasing urbanization process is accompanied by the growing usage of vehicles, which leads to a significant increase in traffic congestion in cities around the world[1–4]. The price of congestion is the enormous time spent on roads[5–8], as well as increasing fuel consumption, air pollution, and Carbone Dioxide emission[6–8, 9] This cost is increasing significantly with urban population growth. Current technological development gave hope that autonomous cars will solve congestion problems as they were expected to reduce the number of private cars by increasing car-sharing. Recent studies, however, show that this is not the case[10–16].

There exists extensive work in various disciplines, e.g. urban planning[2], traffic[17–19], complexity and networks[20–26] that aims at reducing traffic congestion generally, and in urban areas in particular. There are two main approaches to address this issue of reducing congestion: the first is to increase road supply and the second is to reduce the demand for car usage[3,27–29]. The first approach focuses on developing traffic-management solutions and car-sharing systems, while the second usually focuses on implementing taxes or fees on road usage[9,30,31]. Recent studies that considered these two methods[9,32–36] emphasize the urgent need for identification of congestion sources, i.e. the jam bottleneck.

Thus, the objective of this work is to develop and propose a new framework to identify and prioritize traffic bottlenecks, based on big data (retrieved in real time). This methodology can be implemented in planning transportation systems and reduce urban traffic congestion, by combining both of the above approaches. In other words, increasing road supply by means of traffic planning and management systems and reducing the demand by means of a dynamic road-pricing tool.

While existing traffic–management solutions that optimize traffic lights address each intersection individually and use bottom-up solutions such as synchronization and slotting to mitigate local congestion, our method considers simultaneously the entire urban road network and takes into account all the intersections when evaluating jam bottlenecks and different solutions. Currently, there is a lacuna in providing an approach that prioritizes specific bottlenecks over the others in order to optimize the entire road network as well as to provide a dynamic road pricing that charges each vehicle according to its unique effect on the entire system. Our approach provides a holistic framework towards globally optimizing the entire road-network traffic flow, as explained by Hamilton: "When a holistic view of traffic management is taken, individual junction efficiencies can suffer to improve the state of the network as a whole... A strategic view of the entire urban network, with improved detection and communication technologies, is required to enter the next evolution of urban traffic control."[33]

Our present work extends the work presented by Li et al[37] who developed a method to identify traffic jams bottlenecks based on the percolation process while using big data, retrieved in real-time, of traffic speeds. We, however, suggest identifying traffic bottlenecks based on the definition: "The main feature of a bottleneck is that its downstream is in free flow and its upstream is jammed"[38].

Thus, our novel method aims to both identify and prioritize traffic bottlenecks based on big-data of traffic speeds. It is based on the idea that if a bottleneck causes its upstream to be congested, it must be congested prior to it. Hence, for the definition of a bottleneck, time is as important as space.

**Methodology: Identification of traffic bottlenecks**

To identify traffic bottlenecks, similar to[37], we converted three datasets of two urban areas (London, Tel Aviv, and the center of Tel Aviv – without the Ayalon Highway) to dynamic, directed traffic networks where each node represents a junction, and each link represents a street segment between two junctions. The direction of the links represents the allowed traffic on that street segment, and the weight ($W_{ij}(t)$) of the links was defined as:

$$(1) \quad W_{ij}(t) = \frac{U_{ij}(t)}{U_{ij}(P95)}$$

where $U_{ij}(t)$ is the speed in road $ij$ at time $t$ and $U_{ij}(P95)$ is the 95 percentile of all the measured speeds in the studied week. It represents the temporal traffic relative speed availability between 0-1, where 0 stands for heavy traffic (no movement at all) and 1 stands for top speed in this road. There are many approaches to address traffic stream and evaluate urban flow (see a comprehensive review in[45]). In this paper, we follow[45] who suggested a generalized car-following model that bridges microscopic and macroscopic models:

$$(2) \quad U^{1-m} = U_f^{1-m}\left[1 - \left(\frac{k}{k_j}\right)^{l-1}\right].$$

Here, $U$ represents the speed, $U_f$ represents the free-flow speed, $k$ represents the density, and $k_j$ represents jam density. For each street segment, we extracted $U_f$ as the 95% percentile of the maximal measured speed. For each street and for each measurement, we calculated $k$ (based on eq. 2) and q ($q$ is the $Flow = k * U$). We defined a street segment as currently jammed if its speed is below its maximal flow speed. The maximal flow speed of each street segment $U_{ij}(t)_{q_{max}}$ is then inserted to eq. 1 to define its speed availability threshold $W_{jam}$. Lastly, we compared every temporal $W_{ij}(t)$ with $W_{jam}$ to determine if the street segment is currently jammed.

This approach allows us to analyze traffic flow based on the measured speed alone, while eq. 2 allows extracting the temporal density from the temporal speed and maximal speed. However, the presented methodology can be applied to various traffic models (including microscopic models), or even on real density data, collected in real-time. Yet, as will be shown in this paper, even a general model (such as the one we present) can improve our understanding of urban traffic dynamics and the way to manage the system.

Next, we construct for a given time $t$ a new dynamic weighted network, where $W'_{ij}(t)$ is the cumulative time each link has been considered as jammed at $t$ (see figure 1) and used the following process to create tree-shaped clusters of jammed links:

1. At each time $t$, we identify the links with the highest weight $W'$ (i.e. that has been jammed the longest time) and define them as potential trunks of a jam-tree (JT). Next, we identify the branches of the JT by adding links or other trunks, connected to each trunk, with $W' \leq W'_{trunk}$. By doing so, we identify links that became jammed no more than a predefined parameter $\theta$, in this case – defined as 2 measurement units, after the trunk. The value of $\theta$ is only used to limit the connections of new branches to a JT; in other words, it reflects the maximal duration that a jammed street segment is considered as the cause for the jam in its upstream. High values of $\theta$ allow a street segment to connect to its downstream longer times after its downstream became jammed. This leads to larger JTs on one hand but reduces the probability of causality on the other. In other words, in our analysis, if a street segment became jammed no more than 30 minutes after its trunk we can assume that the traffic load in these links resulted from the trunk of the JT. To test this assumption, we compared the result of the analyses of the real data to those of a control random model. The results of this comparison present a qualitative difference, which strengthens our assumption of causality (see figure S1 for elaboration). By using this definition, we consider the trunk as a bottleneck of the JT. Note, that we chose $\theta = 2$ as our datasets had 15 minutes time-intervals and thus, our analysis considered the macro-dynamics of urban traffic. For other datasets with higher resolution of shorter time intervals, different values of $\theta$ can be used.

2. We continue assigning connected links to these JTs in the same iterative process until no more connected links (roads) with $W' \leq \theta$ for the last added branches are found.
3. We start again at stage 1, but now we look for the link with the highest weight $W'$, that has not been assigned to an existing (JT).
4. We continue this process until there are no more jammed links that are not assigned to any JTs.

The resulted clusters represent JTs and the time each of their links was loaded. Examples of JTs are shown in figure 1.

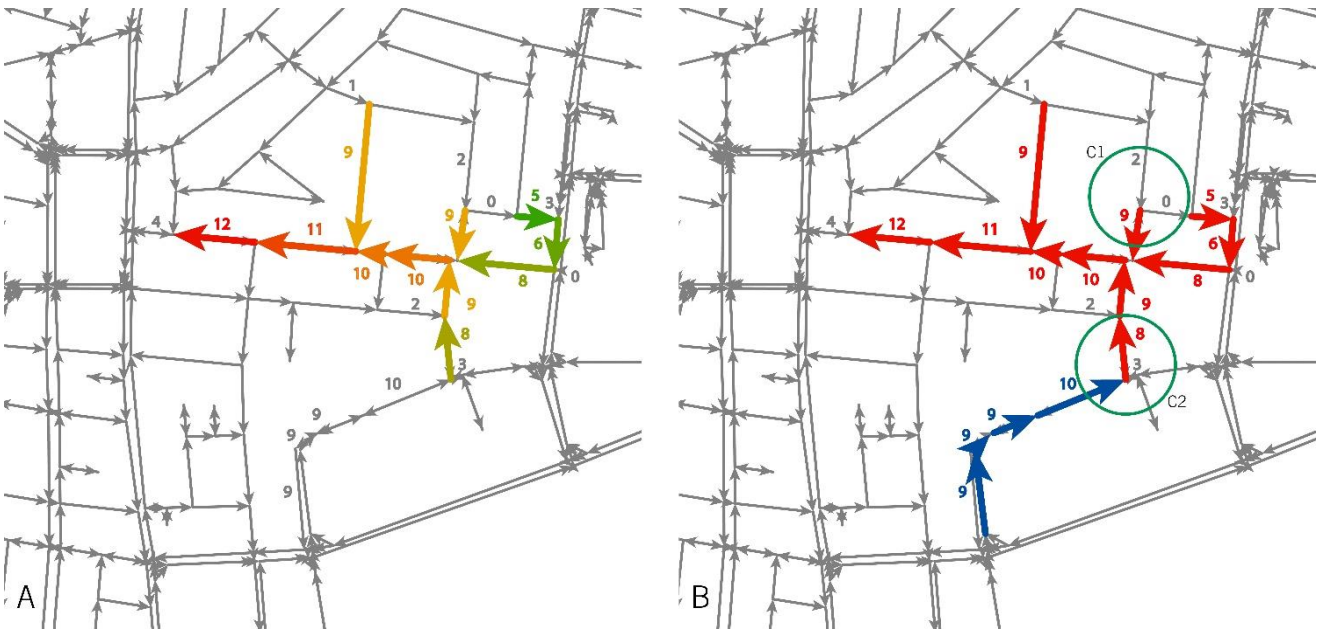

*Figure 1: Clusters of JTs. (A) All the colored streets are part of one JT where the red street represents its trunk: its duration (12 successive measurements that represent 3 hours) is the longest, which indicates it was the first street with traffic load in this JT. (B) Two JTs (represented by red and blue colors). The red JT does not include the street that has been loaded for 2 measurements, as the time gap between this street and its adjacent one is larger than the pre-defined threshold θ (see the upper green circle). The blue JT cannot be considered as part of the red JT, as the duration of its trunk is longer than that of its adjacent street in the red JT (see the lower green circle). When a bottleneck is released but the JT that follows it remains jammed, the next street segment with the longest duration becomes the new trunk of the JT and carries the cost of the remaining branches of the JT.*

While some jams can last many hours, their economic cost might be marginal, if, for example, they occur in peripheral small streets. To assign prioritization for jams, we measure their cost in human hours (HH). In order to calculate the cost at different times of the JTs and the links they include, we introduce the following formulas.

The cost of a link $C_{ij}(t)$ is calculated for every measurement unit -15 minutes in this case, relative to its cost $U_f$, free flow speed. This measurement unit demonstrates the meso-dynamics of the urban traffic. Indeed, using shorter periods of time will allow following the micro-dynamics of urban traffic. This cost represents the time it takes to cross a road (link) in comparison to the time it takes to cross this road in a maximal flow (calculated for each link, based on eq. 2), multiplied by the number of drivers who crossed the endpoint of this link at a specific time:

$$(3)\quad C_{ij}(t) = dist_{ij} * \left(\frac{1}{u_{ij}(t)} - \frac{1}{u_{q_{max_{ij}}}}\right) * \frac{q_{ij}(t)*l_{ij}}{\frac{60}{T}}$$

Here, $dist_{ij}$ is the length of the link in km, $q_{ij}(t)$ is the current flow on the link, $u_{ij}(t)$ is the current speed on the link, $u_{qo_{ij}}$ is the speed when the flow is optimal, $T$ represents a measurement unit which corresponds to 15 minutes (in the present case) and $l_{ij}$ is the number of lanes in the link (i.e. the number of lanes in each street segment of the JT).

The momentary cost of a JT represents the sum of the costs (eq. 3) of all the links that are included in it at a specific measured time:

$$(4)\quad \boldsymbol{MomentaryCost}(t)_{JT} = \sum_{b_{ij}}^{n}(C_{ij}(t))$$

And the cumulative cost of a JT is the cost of the JT from the moment it was created until the time (t) which is calculated as:

$$(5)\quad CumulativeCost(t)_{JT} = \sum_{b_{ij}}\left(\sum_{t_I \leq t}^{t} C_{ij}(t_I)\right)$$

Here, $b_{ij}$ is a branch (i.e. link) in the JT and $t_I$ is the time each branch $b_{ij}$ was a part of the JT (in 15 minutes units).

Figure 2 demonstrates some examples for different $CumulativeCost(t)_{JT}$, based on real data for London and Tel Aviv.

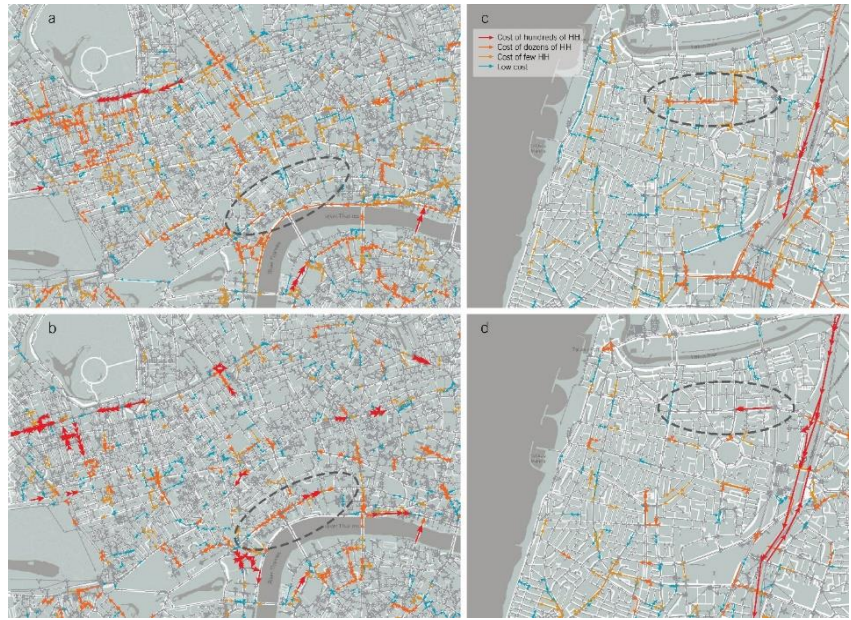

*Figure 2: Graphical representation of the JTs: snapshots of JTs in (a,b) London center and (c,d) Tel Aviv and their $CumulativeCost(t)_{JT}$ during morning rush hours (up) and evening (bottom). While some traffic jams appear in both morning and afternoon rush hours (e.g., Marylebone street in London, or the Ayalon Highway*

*in Tel Aviv), Others are jammed in only one of the rush hours. For example, Victoria Embankment road between Black friars Bridge and Waterloo Bridge (see black circle) is heavily jammed only in the morning rush-hour snapshot; and Pinkas St. in Tel Aviv (see blue circle) is jammed only in the evening rush hours snapshot).*

Lastly, to follow the spatio-temporal dynamics of the system, we combine all the different JTs that had the same street as their trunk throughout the entire examined week and referred to them as Repetitive Jam Trees (RJT). The cumulative cost of the RJTs represents the sum of all the JTs they contain at a specific time window (e.g., day or week):

$$(6) \quad TotalCost_{RJT} = \sum TotalCost_{JT}$$

Equations (4)-(6) allow calculating not only the cost of each JT from the moment it became jammed until it was dissolved but also its dynamics and temporal costs at different times (see figure 2).

**Results: Spatio-temporal dynamics of traffic bottlenecks – macro and meso-dynamics**

We present the results of our analysis through two forms of order: scaling characteristics versus local meso dynamics. We found, that while the data corresponds to scaling laws in a macro resolution[44] when zooming into the spatio-temporal behavior of the bottlenecks and their trees, they vary both in their location and in the time they occur.

<u>Macro-stability</u>

Previous work found universal laws in urban traffic congestion[39–43]. Some studies even identified a high degree of regularity in the measured speed of the street segments[44]. Others focused on the time evolution of urban congestion[42], but not through the analysis of bottlenecks. At large scales, traffic dynamics and jams have been found predictable[44] and their weights follow power-law distributions[42]. With this in mind, we analyzed the bottlenecks' dynamic at the macro-scale to find if they present such regularities as well. We analyzed three temporal scales: the largest scale is a work-week (that includes all the examined working days), the intermediate scale is a 24-hour day, and the microscale is the different hours of the day.

*Large scale – a 5-day work-week:* We explored the behavior of the traffic jams of the different datasets throughout the entire work-week and examined several attributes of the systems: the duration of the jams, their size (in terms of the number of road segments), and their cost (in HH units). For all datasets, the PDFs present, well approximated, power-law distributions[1].

It can be seen that the distributions present similar behavior for all three datasets. This implies that despite the different infrastructure and transportation facilities in these two cities, there may be common characteristics in London and Tel Aviv in terms of their traffic macro-dynamics. While these similarities concern the scaling of probabilities of having a bottleneck of a given size, they do not provide an answer to the question "do the same bottlenecks repeat in time and space?" To try and answer this question, we analyzed the intermediate and short time scales as presented next.

---

[1] Throughout our analysis, we used the maximum likelihood estimator's method[46,47] to test the fit of our data to different heavy-tailed distributions (see Fig. S1 a). In some cases, to allow a comparison between the different distributions, we present a power-law fit even though the data may fit other distributions as well. see figure S2 for further elaboration.

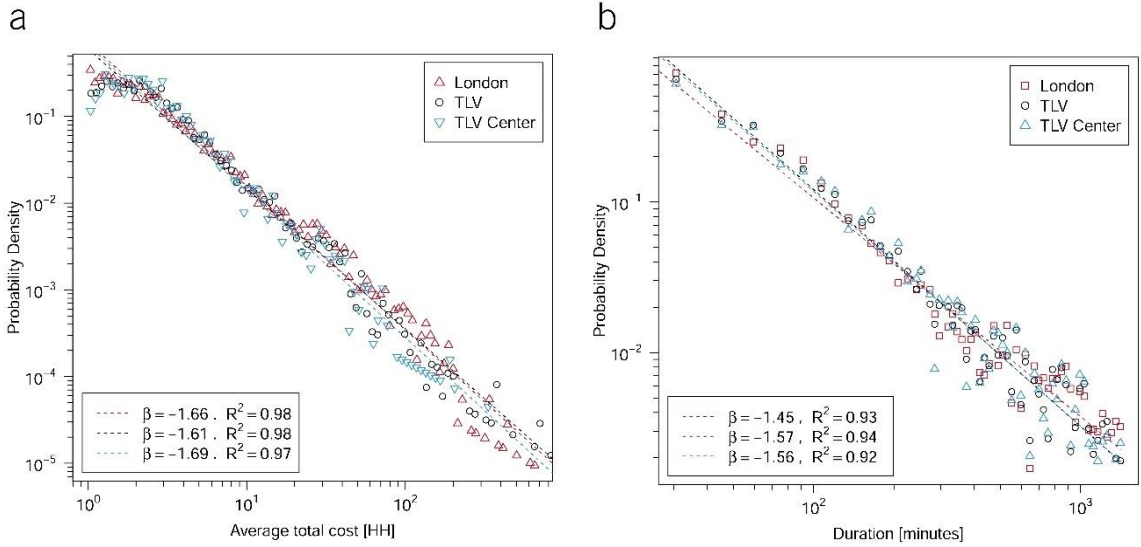

*Figure 3: Analysis of the PDFs of the RJTs in London, Tel Aviv, and Tel Aviv Center (based on the data of all 5 days).* **a.** PDF of average cost, $Ave.Cost_{RJT} = \frac{TotalCost_{RJT}}{N}$, where N represents the number of JTs in the SJT; **b.** PDF of duration (in minutes).

*Intermediate scale – Days:* We examined the PDF of $Max.Cost(t)_{RJT}$ for each dataset separately in each of the examined days (figure S3a-c) and found similar results. The data of London and Tel Aviv (including the Ayalon Highway) present a better fit to a power-law function than the data of Tel Aviv center (without the Ayalon highway). This may be, due to the fact, that the London dataset includes main roads (such as A501) which might have a similar effect on traffic dynamics as the Ayalon Highway. It is worth noting the resemblance between different days on each dataset.

*Short scale – hours:* Lastly, we examined the dynamics of the RJTs in the different datasets and studied their behavior at different hours, days, and cities. By constraining the $TotalCost_{RJT}$ eq. 6 to different time spans, we were able to evaluate their cost at different hours. For that, summing the right-hand side of eq. 6 between $t_1$ and $t_2$, where $t_1$ and $t_2$ correspond to the earlier and later times that define the examined time window. All three datasets present PDFs that correspond to power-law distributions with exponents between 1.8-2.4, where the morning rush-hours present slightly higher values than the afternoon rush-hours in all these cases (figure S3d-f). While the Ayalon Highway introduces costs that reach 1000 HH per hour in both morning and afternoon rush-hours, in London the maximal cost per hour reaches 100 HH in the morning rush-hours, and in Tel Aviv Center, these values are even lower and reach 40HH per hour in the morning rush-hours correspondingly.

We found that the characteristics of the macro-stability indeed align with previous findings, which may be explained by the fact that while different cities have different physical constraints, historical development, and socio-economic processes, urban road networks were developed based on similar principles based on parameters of demand (urban travel) and supply (road infrastructure)[4].

The results of the analysis of all three time scales show that the probability of having a traffic jam of a given cost is scale-free for all cities and in all time spans. Such PDFs can be useful for forecasting the existence of costly traffic jams (above a certain threshold) and the volume of their costs at different time scales. However, the values of these jams (in HH units) and the exponents that govern their decrease with size, depend on spatio-temporal features and represent the different attributes of the different areas. These attributes can relate to the morphology of the street network[48] or other factors such as different types of transportation methods available in each location, working hours norms, etc. Furthermore, while the distributions remain similar on different days in the same city, it is important to study, as we wish here, whether and how much the roads involved in the jam trees remain the same on different days. With this question in mind, we analyzed the meso-dynamics of the traffic networks.

Meso-dynamics

While the bottlenecks cost appears at all scales in different time windows and the traffic load seems regular both in space and time, when zooming into the meso-dynamics of the traffic jams, we unveil local characteristics that reflect the shift in the location of bottlenecks' location over time. These findings reinforce the need to develop new planning tools for urban transportation, that are based on big data.

We analyzed the repetition of bottlenecks on different days and found that most of the bottlenecks are irregular and the same bottlenecks usually do not repeat daily (4a-c). In all three datasets, close to 60% of all bottlenecks appear only in one day of the week. About 20% appear in two days and less than 10% of the bottlenecks with the same level of cost, appear in three days. Even when ignoring their cost levels (see S3), the number of bottlenecks that appear once or twice exceeds 60%. Thus, we see that heavy traffic jams do not repeat daily. Although in all three datasets the bottlenecks with the heaviest cost appear slightly more frequent (the percentage of bottlenecks with cost higher than 100 HH increases for bottlenecks that repeated in 4 or 5 of the days), there exist also many heavy bottlenecks that occurred only once or twice during the examined week. Thus, most of the bottlenecks are not predictable. These results also hold when assuming different thresholds for the bottlenecks' $TotalCost_{RJT}$ (see figure S4). We also compared the above results to the analysis of the bottlenecks' duration (in terms of hours) and found similar behavior. i.e. bottlenecks that lasted longer tend to repeat slightly more frequently than the shorter ones (see figure S4). In other words, the heaviest bottlenecks (in terms of their duration and cost in HH units) are slightly more predictable than the short and less costly ones that occur on different days and locations. Nevertheless, as seen in figure 4, the number of bottlenecks that repeated 4-5 days is only 10-15% of all bottlenecks, where the heavy-loaded ones occupy less than 10% of them. These results are supported by the correlations between the occurrence of bottlenecks in space in time which show similar results (see figures S5 and S6 for elaboration).

We also examine the variation of the $TotalCost(t)_{RJT}$ of the heaviest repeated bottlenecks on different days and at different hours and found that the hourly cost of these bottlenecks in all three datasets changes significantly in different days and hours in both size and rank (see figure 4d-f). These variations are more moderate in London, where the top 5 heaviest bottlenecks remain at the highest ranks (even though their cost significantly changes). In Tel Aviv, the situation is similar in terms of ranks, but the costs vary more significantly. This may be explained by the impact of the Ayalon Highway on the system, as explained earlier. When excluding the Ayalon Highway, Tel Aviv Center presents instability in terms of its bottlenecks' ranks and costs. This holds for all the three examined time spans (see figure S7).

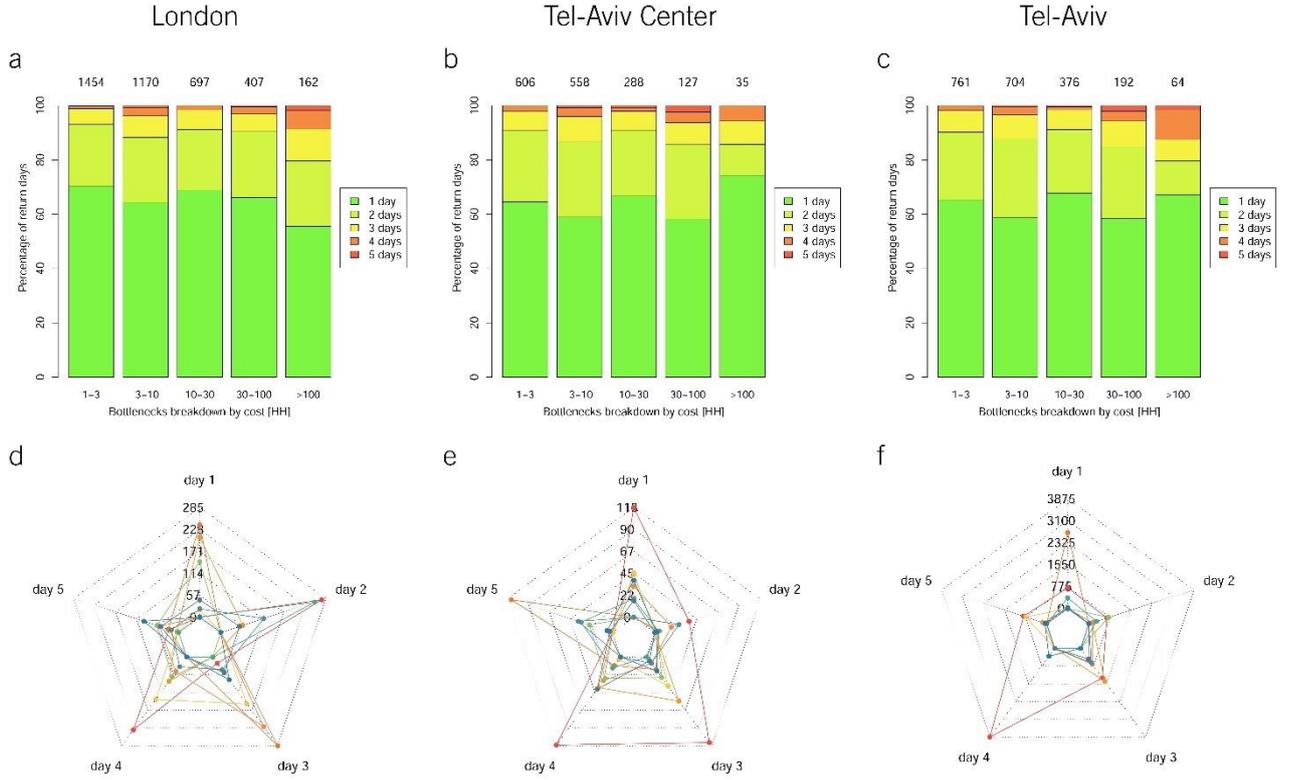

*Figure 4: Top: the repetition of bottlenecks in a. London, b. Tel Aviv Center, and c. Tel Aviv. The X-axis represents the $TotalCost_{RJT}$ of the bottlenecks and the Y-axis represents the percentage of the bottlenecks with different $TotalCost_{RJT}$ repeated during the measured week. Bottom: the changes in the average $TotalCost(t)_{RJT}$ (eq. 6) per hour (in HH units) for the 10 heaviest bottlenecks in d. London, e. Tel Aviv Center, and f. Tel Aviv in 5 days, during $\Delta t$=4 hours from $t_1$=5pm to $t_2$=9pm. These values have been obtained from eq. 5 by summing the right-hand side between $t_1$ and $t_2$.*

Next, we examined the dynamics of the bottlenecks that affected the jammed trees. For that, we analyzed (for each street that was part of any of the traffic jams during our examination time window) the number of different bottlenecks it was connected to. Figure 5 (a-c) shows for each of these streets (in all three datasets), the number of different bottlenecks they were connected to during the 5 days (x-axis), the median distance between these bottlenecks (y-axis), and their relative $TotalCost_{RJT}$ (colored symbols). These results show that jammed streets are connected to a different number of bottlenecks (ranging between 1-22) regardless of their cost (see figure S8 for elaboration on the spatial dynamics of such JTS). However, the bottlenecks in London and Tel Aviv Center are located relatively in proximity to each other (their median distance is less than 1 KM) while the bottlenecks in Tel Aviv are spread over a wider area (up to 2.4 KM), which can be explained by the length of the Ayalon Highway (see above). This means, that while the traffic congestion can be associated with a specific area in the city, and even with some specific streets[44], the location of the bottlenecks that causes the congestion changes constantly on different days and hours. This result suggests that constant, pre-defined solutions for traffic reduction cannot be the only solution to manage urban traffic congestion and strengthen the necessity of using big data in order to improve urban traffic flow. Based on our proposed method, when a trunk is dissolved, the tree can continue to exist with another branch acting as its trunk. To further validate our results, we follow the distribution of the $Max.Cost(t)_{RJT}$ (maximal measured costs for eq. 5) of each tree between all the branches that acted (at any time) as its trunk. Figure 5 (d-f) shows that in more than 80 % of the cases where a tree has more than one trunk during its entire duration, there is a single dominant trunk that holds more than 80% of the tree $CumulativeCost(t)_{JT}$ (eq. 5). In other words, in most cases, there is only one trunk that is responsible for the JT. We also calculated the same frequencies for the duration of the different trunks and found similar results (see figure S9 and S10).

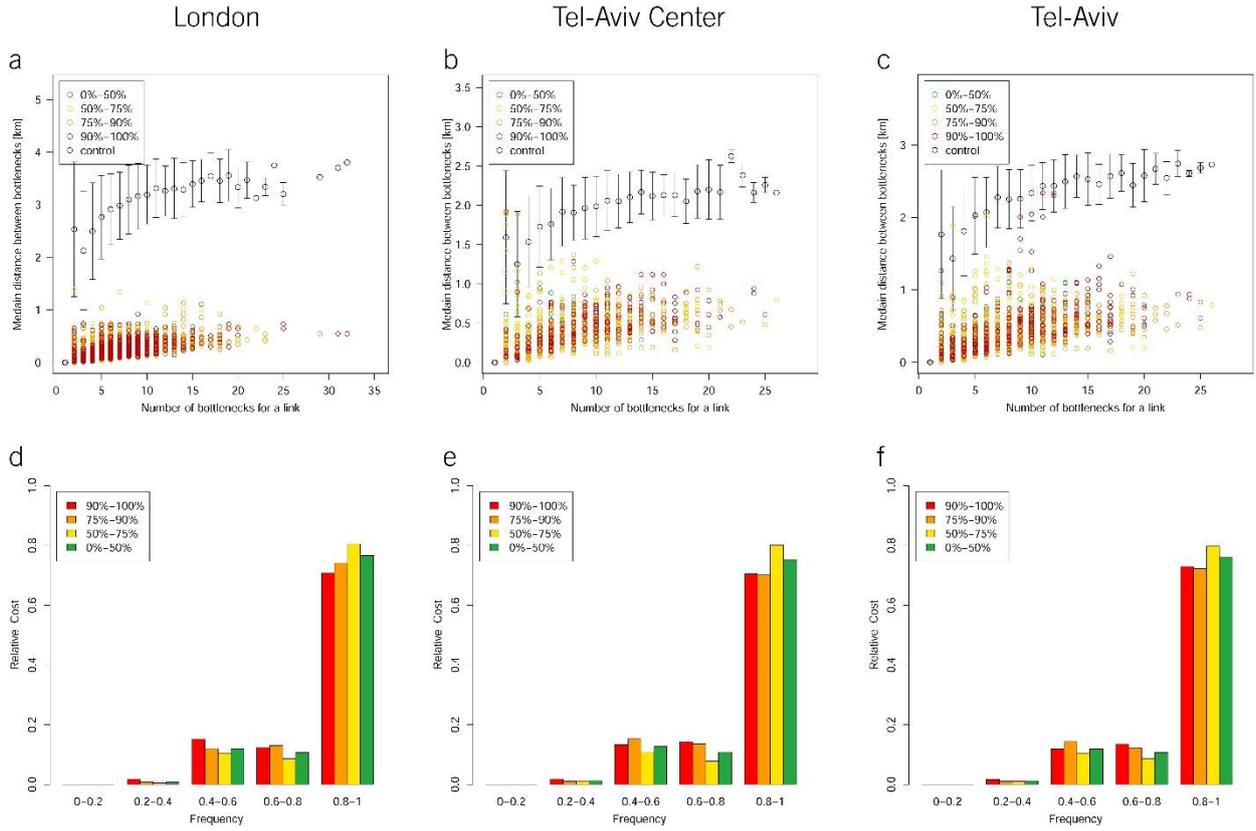

*Figure 5: Top: the streets that were included in any of the JT during the examined week. The X-axis represents the number of different bottlenecks each street was connected to, the Y-axis represents the median distance between these bottlenecks, and the colors represent the relative percentage $TotalCost_{RJT}$ (eq. 6) of each street, in comparison to the other streets in the dataset. The control group is calculated by allocating a number of random bottlenecks to each street as the number of different bottlenecks it was connected to. a. London, b. Tel Aviv Center, and c. Tel Aviv. Bottom: the distribution of the frequency of the relative $CumulativeCost(t)_{JT}$ (eq. 5), i.e, for each tree, the value of the trunk with the highest $CumulativeCost(t)_{JT}$, divided by the sum of the $CumulativeCost(t)_{JT}$ of all the branches that acted as trunks for this tree are presented for d. London, e. Tel Aviv Center, and f. Tel Aviv.*

**Applications – a data-driven example**

Our results show that urban traffic bottlenecks are highly nonrepetitive in terms of their spatio-temporal occurrence, i.e. in most cases, they frequently do not occur in the same location at different times. This finding emphasizes the need to analyze data and identify bottlenecks while accounting for the entire street network (as the method in the current study). Figure 6 presents HaShalom Interchange as an example of the significance of accounting for the entire road network and not analyzing traffic jams at the local level (single or a few adjacent junctions). This area connects several main roads as shown in figure 6. In our analysis, we focus on Ayalon South Highway (B, C, and E), Namir North (F), Kaplan West (G), Giv'at HaTahmoshet West (A), and HaSHalom West (D).

We present an analysis of the traffic congestion in this area, based on the real data, collected on March 21st, 2018 during the morning rush hours. When looking at the traffic congestion at 8:45 all the street segments (with the exception of E – Ayalon South) are jammed. When analyzing the traffic congestions in this area locally, one can assume that road segment G (Kaplan West) may be the reason for the congestion that includes A, B, C, D, and I, as these roads connect to G. However, when analyzing the evolution of the JTs we found that G became jammed only at 8:00, while A and D were the first road segments to become jammed at 6:45, followed by B, and C (both at 7:15). Road segment I was naturally not affected by A, but it also became jammed at 6:45 (an hour and 15 minutes before G). Further investigation reveals that A was not jammed due to the jams in F or L as well, as these roads became jammed only at 7:45 an hour and 15 minutes after A became jammed (see figure S9).

When converting this data into $CumulativeCost(t)_{JT}$ (in HH units, figure S11), from the time each road became jammed until 8:45 am, we found the cost of G was 2HH, while the costs of A and I were 23HH and 157HH respectively. The costs of F and L were 25HH and 26HH respectively. Looking at these costs at the time of the examination, one might assume it is more important to address the jams in I, however, the cost of the JT with A acting as its bottleneck (i.e. A, B, C, and D) was much higher, that is 284HH, suggesting it is more important to address the jam in A first. In fact, B and C are two lanes of the Ayalon Highway that lead to the HaShalom interchange. Thus, it also makes sense that a jam that affects this major highway causes larger damage than a local urban jam, and should be addressed with higher priority.

Although the above example is general and accounts for 15 minutes intervals, it demonstrates that by solving the traffic jams based on the analysis of individual junctions, one can miss the real bottlenecks that have the most significant effect on the overall road network. Naturally, the ability to prioritize the impact of different JTs in the city is not limited only to JTs that reach the same junction but can account for all JTs located in different areas of the city as well.

.

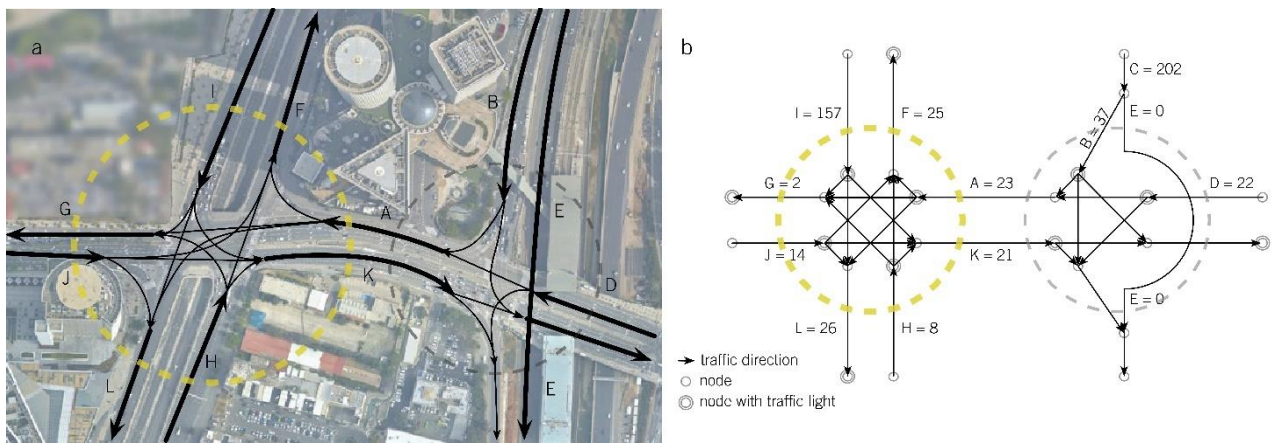

*Figure 6: HaShalom Interchange on March 21$^{st}$, 2018 at 8:45 am. a: Aerial photo of the roads; b. topological representation of the roads network, where the nodes represent junctions (with or without traffic lights), the links represent the road segments, and the weight of the links represents their $CumulativeCost(t)_{JT}$ (in HH units) from the time they became jammed until 8:45 am.*

**Conclusions**

We developed an innovative methodology for identification and prioritization of traffic jams bottlenecks based on big data retrieved in real-time. The online identification of the bottlenecks is based on the idea that a bottleneck must be congested before its upstream, thus considers time as important as space. Their prioritization is based on the analysis of the global urban road network, rather than on local or adjacent junctions (as commonly used in available technologies). We showed that although some universal power-laws distributions that appear daily, govern the macroscopic spatio-temporal behavior of traffic jams, there are also unique behaviors that indicate that local attributes affect traffic dynamics as the same traffic bottlenecks usually do not reappear on different days. In other words, the macro-stability, presented by the scaling characteristics of the traffic bottlenecks that represent the seeming regularity of traffic load both in time and space, overshadows the existence of meso-dynamics, where the bottlenecks that create these JT loads, change their location in time and space. This means that in order to manage traffic jams in different locations and at different times and determine priorities, there is a need to implement unique solutions that track traffic and evaluates the relative effect of each bottleneck on the entire road network. Our method can extend existing systems where big data is used to identify traffic jams. This method can assist in identifying and prioritizing the bottlenecks based on their cost (for example - in human-hours units). This will, of course, require more accurate and detailed datasets, However, in the era of smartification of the cities, that can be obtained by IoT technologies (i.e., sensors) or ICT (i.e., real-time navigation apps). By using such data, our framework might be used to develop new planning tools that allow increasing road supply by means of prioritizing the improvement of specific bottlenecks over the others. This can be done based on the bottlenecks' cost (as shown in this work) or based on other considerations such as evacuation during extreme events, helping emergency vehicles reach their destination faster, etc. Additionally, such a system can help reducing the demand by means of a dynamic road-pricing tool that is based on the relative cost each bottleneck causes. As this method is based on real-data, it will be able to constantly feedback itself and better control and mitigate traffic jams. Implementation of this method may be a part of real-life systems, leading to a breakthrough in dealing with urban traffic around the world.

**Data**

The results of our framework and analysis are demonstrated for three urban areas: London center, Tel Aviv center (including the Ayalon highway – the main road that crosses the city from North to South), and Tel Aviv Center (without the Ayalon highway).

These cases were chosen as they represent cities of different scales (London center is 2.5 times larger than Tel Aviv center); different public transportation systems and different regulations that influence the driving behavior; and the exclusion of the Ayalon highway from the data of Tel Aviv also allows us to explore the effect of a local highway on the local transportation characteristics.

We collected from Google Directions API the speeds of 8,857 road sections in London center and 2,950 road sections in Tel Aviv (2324 of which are of Tel Aviv Center), every 15 minutes over a week's time. The data for London center was collected between the dates 21-27/3/2018 and the data for Tel Aviv center has been collected between the dates 12-18/2/2017). We developed an algorithm that considers additional road segments (for which we did not have data) based on interpolating the data collected for their adjacent road segments and ended up with 18,050 road sections in London, 5,425 road sections in Tel Aviv, and 3871 road sections in Tel Aviv Center.

For each case, we analyzed the data of 5 working days only (Mon-Fri in London and Sun-Thursday in Tel Aviv). This is because the results indicate that the dynamics of these systems are significantly different during the weekends.

**Data Availability**

For contractual reasons, we cannot make the empirical data from Google Direction available. However, all data from our analysis are available at GitHub:

https://github.com/nimrodSerokTAU/bottlenecks-prioritization

**Code Availability**

The code of our analysis is also available at GitHub:
https://github.com/nimrodSerokTAU/BottlenecksPrioritization

## Acknowledgments

Financial support for this study was provided by a grant from the Center for Innovative Transportation for supporting this research.

## Author contributions

Conceived and design the methodology: N.S., S.H., E.B.L. Executed the code: N.S. Analyzed the data: N.S., S.H., E.B.L. Wrote the paper: N.S., S.H., E.B.L.

## Competing interests

The authors declare no competing interests.

## Materials & Correspondence

Efrat Blumenfeld Lieberthal efratbl@tauex.tau.ac.il